\documentclass[]{aa} 
 
\usepackage{graphicx, times, amsfonts}

\newcommand{\less}{\raisebox{-1.1mm}{$\stackrel{<}{\sim}$}} 
\newcommand{\more}{\raisebox{-1.1mm}{$\stackrel{>}{\sim}$}}

\newcommand{\OG}{{\sc ogle}} 
 
\newcommand{\tc}{Population {\sc ii} Cepheids} 
 
\begin{document} 
 


\title{The distance to the Galactic Centre based on Population {\sc ii} Cepheids and RR Lyrae stars\thanks{Based 
on observations collected at the European Southern Observatory, Chile (ESO Programme 079.B-0107).
%
}
}  
 
\author{ 
M.A.T.~Groenewegen 
\inst{1}  
\and 
A.~Udalski
\inst{2}  
\and 
G.~Bono
\inst{3,4}  
}

\institute{ 
Instituut voor Sterrenkunde, Celestijnenlaan 200 D,  
B--3001 Leuven, Belgium \\ \email{groen@ster.kuleuven.be}
\and
Warsaw University Observatory, Aleje Ujazdowskie 4, PL-00-478, Warsaw, Poland
\and
INAF Osservatorio Astronomico di Roma, Via Frascati 33, I-00040 Monte Porzio Catone, Italy
\and
European Southern Observatory, Karl-Schwarzschild-Str. 2, D-85748 Garching 
bei Munchen, Germany
} 
 
\date{received: 2007,  accepted:  2008} 
 
\offprints{Martin Groenewegen} 
 
 
\abstract{The distance to the Galactic Centre (GC) is of importance
for the distance scale in the Universe.  The value derived by
Eisenhauer et al. (2005) of 7.62 $\pm$ 0.32 kpc based on the orbit of
one star around the central black hole is shorter than most other
distance estimates based on a variety of different methods.}
{To establish an independent distance to the GC with high accuracy. 
To this end Population-{\sc ii} Cepheids are used that have been discovered in 
the \OG-{\sc ii} and \OG-{\sc iii} surveys.}
{Thirty-nine Population-{\sc ii} Cepheids have been monitored with the
SOFI infrared camera on 4 nights spanning 14 days, obtaining typically
between 5 and 11 epochs of data. Light curves have been fitted using
the known periods from the OGLE data to determine the mean $K$-band
magnitude with an accuracy of 0.01-0.02 mag. It so happens that 37 RR
Lyrae stars are in the field-of-view of the observations and mean
$K$-band magnitudes are derived for this sample as well.}
{After correction for reddening, the period-luminosity relation of
Population-{\sc ii} Cepheids in the $K$-band is determined, and the
derived slope of $-2.24 \pm 0.14$ is consistent with the value derived
by Matsunaga et al. (2006). Fixing the slope to their more accurate
value results in a zero point, and implies a distance modulus to the
GC of 14.51 $\pm$ 0.12, with an additional systematic uncertainty of
0.07 mag.  Similarly, from the RR Lyrae $K$-band period-luminosity
relation we derive a value of 14.48 $\pm$ 0.17 (random) $\pm$ 0.07 (syst.).
The two independent determinations are averaged to find 
14.50 $\pm$ 0.10 (random) $\pm$ 0.07 (syst.), or 7.94 $\pm$ 0.37 $\pm$ 0.26 kpc.
The absolute magnitude scale of the adopted period-luminosity
relations is tied to an LMC distance modulus of 18.50 $\pm$ 0.07.  }
{}


\maketitle

\section{Introduction} 

The distance to astronomical objects is a crucial parameter, yet often
very difficult to obtain with high precision. The distance to the
Galactic Centre (GC) is of special importance, e.g for dynamics (Oort
constants, determining distances using a rotation model), or for
calibrating standard candles.
The classically accepted value comes from the review by Reid (1993)
and is $R_0$ = 8.0 $\pm$ 0.5 kpc.

Over the last few years the distance to the GC based on the orbit of the
star called S2 around the central black-hole (BH) has caught
attention.  Initially, Eisenhauer et al. (2003) derived a value of
7.94 $\pm$ 0.42 kpc which was revised by Eisenhauer et al. (2005) to
7.62 $\pm$ 0.32 kpc having more epochs of data available.
The neglect of post-Newtonian physics in these analysis may have lead
to an underestimate of the distance by about $0.11 \pm 0.02$ kpc
(Zucker et al. 2006), leading to a current best estimate of 7.73 $\pm$
0.32 kpc (corresponding to a distance modulus (DM) of 14.44 $\pm$ 0.09) 
to the GC based on the BH.

On the other hand, most other recent distance determinations give a
longer distance, more in line with the classical value:
(1) High-amplitude delta-scuti stars give 7.9 $\pm$ 0.3 kpc (McNamara et al. 2000); 
(2) RR Lyrae stars suggest a value of 8.8 $\pm$ 0.3 kpc (Collinge et
    al. 2006), 8.3 $\pm$ 1.0 kpc (Carney et al. 1995) or 8.0 $\pm$
    0.65 kpc (Fernley et al. 1987).
(3) Earlier work on the Red Clump gave a longer distance of 8.4 $\pm$ 0.4 kpc 
(Paczy\'nski \& Stanek 1998), although Nishiyama et al. (2006) derive
7.52 $\pm$ 0.10 (stat) $\pm$ 0.35 (syst) kpc, and Babusiaux \& Gilmore (2005) 
7.7 $\pm$ 0.15 kpc; 
(4) From a comparison of Miras found in the OGLE database in the direction of
the Galactic Bulge (GB) to those in the Magellanic Clouds, Groenewegen \&
Blommaert (2005) find a distance in the range 8.5 to 9.0 kpc, in
agreement with earlier work on Miras (Catchpole et al. 1999); 
(5) Analysis of the Hipparcos proper motions of 220 Cepheids lead to 
$R_0$ = 8.5 $\pm$ 0.5 kpc (Feast \& Whitelock 1997); 
(6) Modelling the observed colour-magnitude diagram in $V,I$ and $J,K$ 
using a population synthesis code, Vanhollebeke et al. (2008) 
derive a distance of 8.60 $\pm$ 0.16 kpc.

With the exception of some Red Clump based distances, the results
obtained by Eisenhauer et al. imply a much shorter distance to the GC
than found by most other methods, and this calls for an independent
investigation of this matter.

In this paper the distance to the GC is determined using \tc\
(hereafter P2C) discovered in the OGLE micro-lensing survey, and for
which the mean $K$-band magnitude will be determined by infrared
monitoring. Comparing to the calibrated P2C period-luminosity (PL)
relation in the $K$-band from Matsunaga et al. (2006; hereafter M06)
then provides the distance, after correction for reddening.

In addition, the mean $K$-band magnitude will be determined for RR
Lyrae stars that are in the field, and compared to the calibrated
$K$-band PL-relation from Sollima et al. (2006). The Matsunaga et
al. and Sollima et al. relations both imply an LMC DM of 18.50 as
detailed in Sect.~5.

In Sect.~2 the sample is discussed, and the observations are presented
in Sect.~3 for the P2C and Sect.~3 for the RR Lyrae.  The results are
discussed in Sect.~5.

\section{The sample of \tc}

Population-{\sc ii} Cepheids are old, low-mass stars. They are the
progeny of hot HB stars that after the exhaustion of core He-burning,
move toward lower effective temperatures (post-early AGB), thus
crossing the Cepheid instability strip. They are systematically
brighter than RR Lyrae stars and have periods ranging from slightly
below one to a few tens of days

Kubiak \& Udalski (2003; hereafter KU) have searched the OGLE-{\sc ii}
database for P2C Cepheids and found 54 objects. 
KU determined a period-luminosity relation in the reddening-free
Wesenheit index (based on $V,I$ photometry) and compared it to P2C
Cepheids in the LMC. The difference in DM at a typical period of $\log
P$ = 0.5 is 3.58.  By assuming a DM for the LMC of 18.5 (50.1 kpc) the
distance to the GC P2C becomes 9.6 kpc (with a substantial error bar
of 1.5 kpc).  This places the objects nominally in the GB
region. Additional evidence is that KU found these objects to be
located in a bar (see their Fig.~2), like the RR Lyrae (Collinge et
al. 2006), a strong indication that the P2C Cepheids indeed are
physically part of the GC region. These properties gave us the
confidence that determining individual distances to these objects
would indeed provide an accurate distance determination to the GC.

\begin{figure}

\begin{minipage}{0.49\textwidth}
\resizebox{\hsize}{!}{\includegraphics{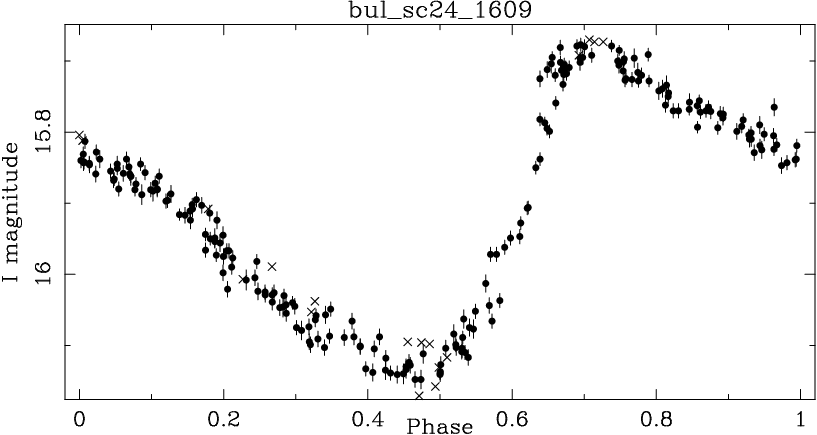}}
\end{minipage}
\begin{minipage}{0.49\textwidth}
\resizebox{\hsize}{!}{\includegraphics{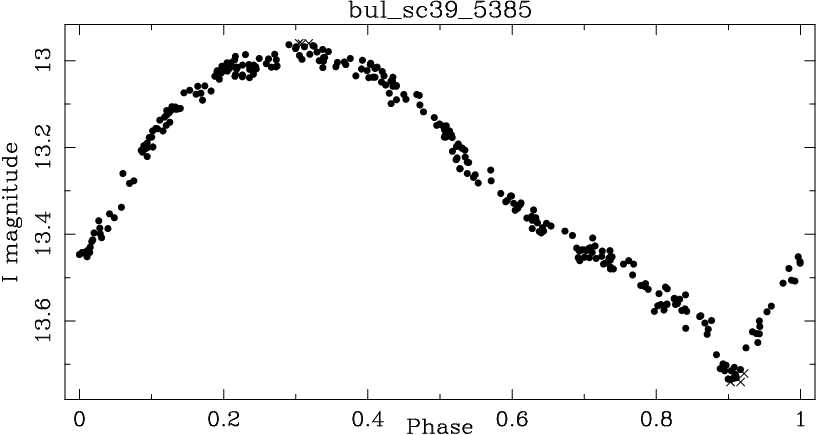}}
\end{minipage}

\caption[]{
Phased OGLE $I$-band light curves of two of the P2C.
The periods of the two stars are 0.76522 (bul\_sc24\_1609) and 9.94431 (bul\_sc39\_5385) days. 
}
\label{Fig-ILC}
\end{figure}

A subset of stars was selected based on the following criteria:
The $I$-band light curves of these stars were inspected and only
``smooth'' ones were retained that show no or little effect of shocks in the
atmosphere (Figure~\ref{Fig-ILC} shows two examples).
The range in Galactic longitude was largely restricted to 
-2.5 \less $l$ \less +2.5 degrees to avoid any additional smearing 
in distance due to the effect of the inclination of the bar.
The entire period range from 0.8 to $\pm$10 days should be covered.
OGLE $I$-band images were inspected for crowding.

In a later stage, it was also possible to extent the sample with the
first results from the ongoing OGLE-{\sc iii} survey and a sample of
70 new P2C in the longitude range -2.5 to +2.5 degrees could be added.  
After inspecting the light curves and $I$-band images of these stars
as well, a final combined sample of 49 stars remained, of which 39
were actually monitored.

The basic properties of the sample are listed in Table~\ref{Tsample}
in the first columns: Identification, R.A., Declination, pulsation
period, galactic coordinates.

\section{The observations} 
 
The observations were carried out with the SOFI infra-red camera on 
the 3.5m NTT on ESO/La Silla in the nights of 2007, June 24, 28, July 3, 8 in visitor mode.

Photometric conditions were excellent on the second and third night
with seeing as low as 0.6\arcsec\ and the telescope was actually defocused.  
Weather conditions were poorer on the first and last night with seeing 
in the range 1.5-2\arcsec\ and cirrus and thin clouds. As the
measurements of the P2C will be relative to the 2MASS objects in the
field this additional extinction did not influence this program.

Typically it was tried to observe the longest period (\more 7 d)
Cepheids at the beginning and end of the night, the intermediate
period ones (\more 3 d) three times per night, and the shorter period
ones were observed one after the other over the entire night. 
In total we obtained 362 epochs of data of 39 P2C.

Images in the $K_{\rm s}$ band (hereafter simply $K$-band) were taken
with the shortest possible on-chip integration time of DIT= 1.2
seconds using a pixel-scale of 0.288\arcsec\ and resulting in a
field-of-view of almost 5$\times$5 arcmin\footnote{The 0.144\arcsec\
pixel scale was unfortunately no longer offered.}. The ``auto-jitter''
observing block (OB) was used with 9-13 exposures. The relative large
number was used to have sufficient redundancy in creating a sky image
in these relatively crowded fields. The $K$-band was chosen for the observations 
as the dispersion in the $PL$-relation is smaller in that band than in $J$ or $H$ (M06).

The data were reduced with the newly released SOFI data reduction
pipeline\footnote{Available at
http://www.eso.org/sci/data-processing/software/pipelines/}. 
The pipeline takes into account the cross-talk, the flat-field 
(dome flats taken with a special OB), bad pixel cleaning,
and images correlation and reconstruction.

\begin{figure}

\begin{minipage}{0.49\textwidth}
\resizebox{\hsize}{!}{\includegraphics{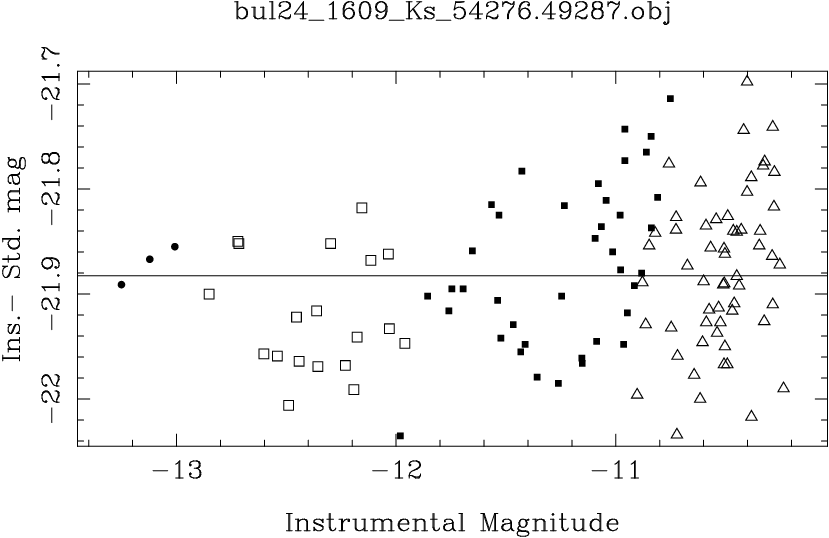}}
\end{minipage}
\begin{minipage}{0.49\textwidth}
\resizebox{\hsize}{!}{\includegraphics{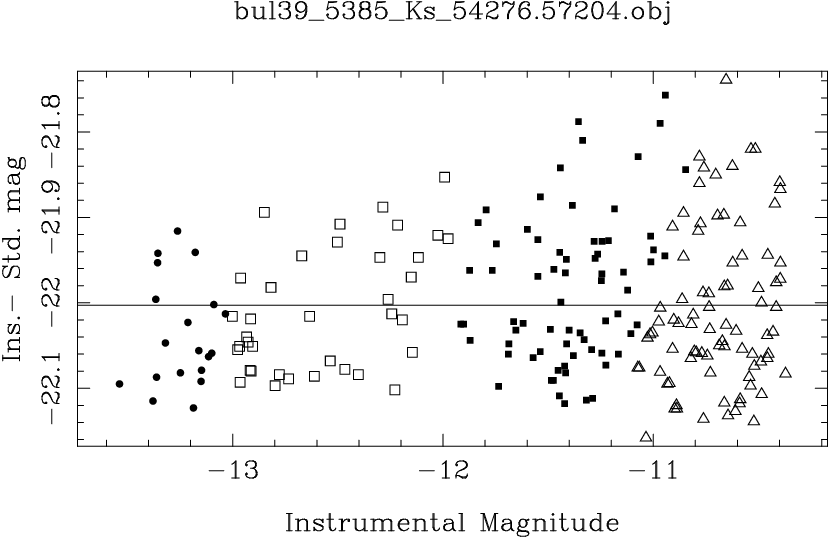}}
\end{minipage}

\caption[]{
The difference between SOFI instrumental magnitude and 2MASS magnitude 
versus instrumental magnitude for a single epoch for two different objects. 
The symbols indicate different 1 magnitude bins in 2MASS magnitude.
%
}
\label{Fig-ZP}
\end{figure}

The reduced images were trimmed to the original 1024 $\times$ 1024 pixel size to eliminate the under exposed edges.
The astrometric solution was done using the WCSTools
suite\footnote{available at
http://tdc-www.harvard.edu/software/wcstools/} matching the stars in
the field against the 2MASS catalog. Typically several tens of 2MASS
objects are available and the rms in the solution typically is 0.2\arcsec.
Source extraction and PSF photometry was done using version 1.3 of DoPhot (Schechter et al. 1993).
A dedicated Fortran program was written to match the 2MASS objects
with the sources in the field, and determine the offset between the
instrumental magnitude and the 2MASS $K_{\rm s}$-band magnitude. The
brightest and faintest 2MASS objects were excluded to avoid problems
of saturation in the SOFI images and poor quality 2MASS data. 
A straight line was fitted to the data, and this determined the offset
to be applied to the SOFI instrumental magnitudes to put them on the
2MASS system (an example is shown in Fig.~\ref{Fig-ZP}).
The typical error in the determination of this offset is 0.006 mag, and
this is added in quadrature to the error in the instrumental magnitude
for each star to give the total error in the observed magnitude.

This error is verified by comparing the magnitudes over the different
epochs for field stars and estimating the precision with which the
mean magnitude can be determined.


\begin{figure*}

\begin{minipage}{0.49\textwidth}
\resizebox{\hsize}{!}{\includegraphics{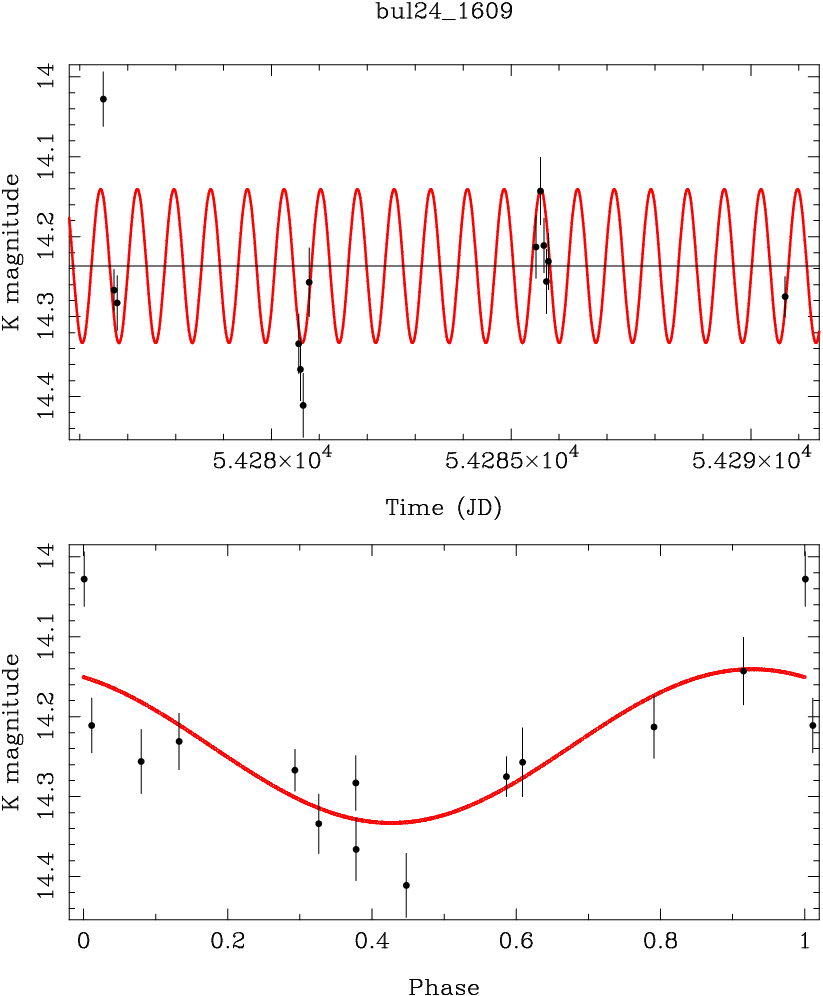}}
\end{minipage}
\begin{minipage}{0.49\textwidth}
\resizebox{\hsize}{!}{\includegraphics{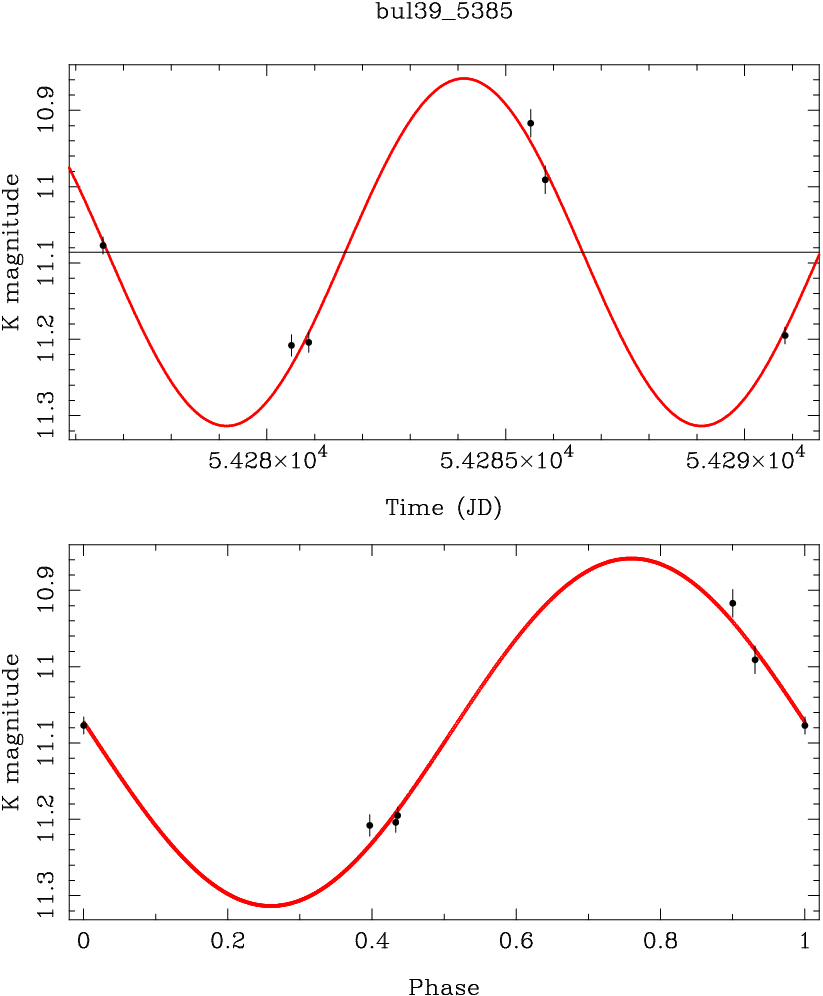}}
\end{minipage}

\caption[]{
The $K$-band light curves of two P2C.
The top panel shows the magnitude versus time, the bottom panel the phased light curve.
}
\label{Fig-KLC}
\end{figure*}

After having determined the photometric offsets, the light curves of
the P2C could be constructed.  The light curves were then fitted
with a sine-function with the known period from the OGLE data, and
this allowed the determination of the mean $K$-magnitude with high
precision.  Figure~\ref{Fig-KLC} shows the (phased) light curve for
two P2C Cepheids with a short and long period in the sample.  The
last three columns of Tab.~\ref{Tsample} list the mean $K$-band
magnitude, the precision in this determination and the number of epochs.

\begin{table*} 
\setlength{\tabcolsep}{1.2mm}
\caption{\label{Tsample} The sample of Cepheids, ordered by period. 
} 
\begin{tabular}{lccrrrcrlccrlccrlccr}\hline \hline
Name             & RA         & DEC          & Period  & $l$ & $b$ & $A_{\rm K}^{(1)}$ & $\sigma$ & $n$ & $A_{\rm K}^{(2)}$ & $A_{\rm K}^{(3)}$ & $\sigma$ & $n$ & $K$ & $\sigma$ & $n$ \\
                 & (J2000)    & (J2000)      & (days)  &     &     &            &          &     &             &             &          &     &     & & & \\
\hline 

bul24\_1609      & 17 53 44.41 & -32 57 14.0 & 0.76522 & 357.44 & -3.56 &  0.284 & 0.018 & 6 &  0.302 &     - &     - &  - &  14.237 & 0.010 & 13 \\

bul214.2\_135    & 17 57 35.12 & -28 26 11.6 & 0.82684 &   1.77 & -2.00 &  0.317 & 0.028 & 5 &  -     & 0.354 & 0.011 &  6 &  13.805 & 0.049 &  5 \\
bul214.6\_53717  & 17 56 15.76 & -28 14 02.3 & 0.83088 &   1.80 & -1.65 &  0.352 & 0.005 & 3 &  -     &     - &     - &  - &  14.118 & 0.057 &  6 \\
bul39\_2239      & 17 56 05.04 & -29 54 51.8 & 0.92213 &   0.33 & -2.46 &  0.275 & 0.026 & 6 &  0.316 & 0.343 & 0.023 &  6 &  13.836 & 0.023 &  6 \\
bul22\_3993      & 17 56 33.03 & -30 36 33.5 & 0.93110 & 359.77 & -2.89 &  0.330 & 0.051 & 6 &  0.329 &     - &     - &  - &  13.583 & 0.013 &  9 \\
bul31\_662       & 18 01 56.92 & -28 55 11.6 & 0.93951 &   1.82 & -3.07 &  0.193 & 0.019 & 6 &  0.217 & 0.237 & 0.021 &  3 &  13.925 & 0.010 &  8 \\
bul43\_351       & 17 35 08.13 & -27 31 25.8 & 1.09776 & 359.97 &  2.71 &  0.492 & 0.045 & 6 &  0.440 &     - &     - &  - &  13.629 & 0.007 & 13 \\
bul163.7\_46320  & 17 51 54.32 & -31 56 39.8 & 1.21123 & 358.12 & -2.71 &  0.433 & 0.078 & 4 &  -     &     - &     - &  - &  12.573 & 0.010 & 11 \\
bul30\_1107      & 18 01 47.33 & -29 07 39.1 & 1.33916 &   1.63 & -3.14 &  0.215 & 0.016 & 4 &  0.229 & 0.233 & 0.015 &  4 &  12.126 & 0.006 & 11 \\
bul215.7\_107864 & 17 58 54.55 & -28 21 30.8 & 1.40015 &   1.98 & -2.22 &  0.282 & 0.035 & 6 &  -     & 0.278 & 0.011 &  4 &  13.088 & 0.009 & 11 \\
bul3\_791        & 17 53 27.82 & -30 19 55.3 & 1.48412 & 359.68 & -2.18 &  0.323 & 0.049 & 5 &  0.347 &     - &     - &  - &  14.255 & 0.022 &  8 \\
bul5\_3719       & 17 50 21.49 & -29 54 27.5 & 1.50105 & 359.70 & -1.39 &  0.310 & 0.023 & 4 &  0.688 &     - &     - &  - &  13.304 & 0.028 &  7 \\
bul32\_2167      & 18 02 56.05 & -28 40 39.2 & 1.50523 &   2.14 & -3.14 &  0.178 & 0.012 & 4 &  0.193 & 0.212 & 0.011 &  5 &  13.351 & 0.008 & 11 \\
bul4\_170        & 17 54 38.21 & -30 10 41.8 & 1.53185 & 359.94 & -2.32 &  0.278 & 0.027 & 5 &  0.311 &     - &     - &  - &  13.027 & 0.010 & 10 \\
bul44\_5324      & 17 49 30.95 & -29 50 58.2 & 1.55140 & 359.66 & -1.20 &  0.313 & 0.039 & 5 &  0.720 &     - &     - &  - &  13.449 & 0.011 &  8 \\
bul188.1\_11087  & 17 59 30.71 & -30 19 59.7 & 1.69825 &   0.33 & -3.31 &  0.271 & 0.045 & 5 &  -     &     - &     - &  - &  12.904 & 0.018 &  9 \\
bul21\_3035      & 17 59 53.52 & -28 56 37.5 & 1.73020 &   1.58 & -2.69 &  0.198 & 0.019 & 5 &  -     & 0.219 & 0.014 &  5 &  13.159 & 0.017 &  9 \\
bul45\_1189      & 18 03 33.61 & -30 01 14.6 & 1.74795 &   1.04 & -3.92 &  0.177 & 0.019 & 6 &  0.220 & 0.169 & 0.009 &  5 &  13.125 & 0.018 &  9 \\
bul23\_611       & 17 58 14.58 & -31 33 23.7 & 1.85470 & 359.13 & -3.68 &  0.312 & 0.009 & 4 &  0.324 &     - &     - &  - &  12.783 & 0.007 & 10 \\
bul34\_4631      & 17 58 01.53 & -28 59 56.6 & 2.33010 &   1.33 & -2.37 &  0.234 & 0.031 & 4 &  0.302 & 0.312 & 0.033 &  5 &  12.690 & 0.024 &  9 \\
bul20\_961       & 17 58 55.93 & -29 10 57.0 & 2.88445 &   1.27 & -2.63 &  0.223 & 0.038 & 6 &  0.233 & 0.219 & 0.004 &  5 &  12.661 & 0.006 &  9 \\
bul41\_3841      & 17 51 42.51 & -32 41 41.3 & 3.20082 & 357.45 & -3.06 &  0.346 & 0.028 & 5 &  0.318 &     - &     - &  - &  12.196 & 0.008 &  9 \\
bul4\_8846       & 17 54 06.61 & -29 16 22.1 & 3.51336 &   0.66 & -1.76 &  0.295 & 0.025 & 5 &  0.311 &     - &     - &  - &  12.449 & 0.012 &  6 \\
bul4\_2323       & 17 54 55.52 & -29 57 31.0 & 3.54254 &   0.16 & -2.26 &  0.271 & 0.022 & 6 &  0.311 & 0.286 & 0.010 &  5 &  13.267 & 0.112 &  4 \\
bul39\_616       & 17 55 12.35 & -30 07 24.1 & 3.65017 &   0.05 & -2.40 &  0.268 & 0.023 & 4 &  0.316 &     - &     - &  - &  12.281 & 0.007 &  8 \\
bul195.4\_144166 & 17 53 40.70 & -28 56 47.0 & 3.82791 &   0.90 & -1.52 &  0.315 & 0.024 & 4 &  -     &     - &     - &  - &  12.101 & 0.014 &  4 \\
bul22\_815       & 17 56 46.47 & -31 07 07.3 & 4.49944 & 359.36 & -3.19 &  0.270 & 0.031 & 5 &  0.329 &     - &     - &  - &  12.415 & 0.035 &  7 \\
bul3\_1755       & 17 53 34.75 & -30 12 39.7 & 4.57875 & 359.80 & -2.14 &  0.318 & 0.046 & 6 &  0.347 &     - &     - &  - &  12.613 & 0.055 &  7 \\
bul38\_3260      & 18 01 32.68 & -29 49 11.5 & 5.53097 &   1.00 & -3.44 &  0.214 & 0.027 & 7 &  0.220 & 0.206 & 0.018 &  4 &  12.175 & 0.121 &  5 \\
bul180.5\_151030 & 17 51 57.55 & -30 21 39.2 & 6.89131 & 359.49 & -1.91 &  0.381 & 0.042 & 5 &  -     &     - &     - &  - &  11.601 & 0.007 &  6 \\
bul20\_3867      & 17 58 55.95 & -28 43 19.5 & 7.13700 &   1.67 & -2.40 &  0.233 & 0.031 & 4 &  0.233 & 0.243 & 0.012 &  5 &  11.639 & 0.007 &  6 \\
bul2\_1657       & 18 04 19.87 & -29 02 44.2 & 7.45600 &   1.97 & -3.59 &  0.174 & 0.007 & 6 &  0.186 & 0.189 & 0.014 &  4 &  11.464 & 0.007 &  6 \\
bul25\_693       & 17 54 24.38 & -33 06 51.0 & 7.73934 & 357.37 & -3.76 &  0.275 & 0.020 & 6 &  0.281 &     - &     - &  - &  11.775 & 0.006 &  6 \\
bul40\_2524      & 17 50 58.92 & -33 08 53.1 & 7.77000 & 356.98 & -3.16 &  0.344 & 0.022 & 6 &  0.353 &     - &     - &  - &  11.999 & 0.006 &  6 \\
bul214.5\_169880 & 17 56 37.75 & -28 01 28.4 & 8.07957 &   2.02 & -1.61 &  0.367 & 0.050 & 5 &  -     & 0.462 & 0.017 &  5 &  12.180 & 0.007 &  5 \\
bul7\_540        & 18 08 44.36 & -32 13 10.6 & 9.52109 & 359.64 & -5.95 &  0.149 & 0.004 & 4 &  0.160 &     - &     - &  - &  11.220 & 0.006 &  6 \\
bul39\_5385      & 17 55 23.12 & -29 31 35.3 & 9.94431 &   0.58 & -2.13 &  0.278 & 0.025 & 4 &  0.316 & 0.323 & 0.006 &  4 &  11.086 & 0.007 &  6 \\
bul334.7\_59737  & 17 40 50.79 & -25 03 52.7 & 11.38492 &  2.74 &  2.94 &  0.456 & 0.036 & 7 &  -     &     - &     - &  - &  14.200 & 0.018 &  5 \\
bul333.2\_25311  & 17 36 34.05 & -27 18 08.3 & 14.89102 &  0.33 &  2.56 &  0.449 & 0.034 & 4 &  -     &     - &     - &  - &  10.856 & 0.008 &  5 \\

\hline 
\end{tabular} 
Notes:

(1) Reddening from Marshall et al. (2006).
(2) Reddening based on Sumi (2004).
(3) Reddening based on Popowski et al. (2003).
\end{table*}

\begin{table*} 
\caption{\label{TsampleRRL} The sample of RR Lyrae stars, ordered by period.
} 
\begin{tabular}{lccrrrcrlccr}\hline \hline
Name           & RA       & DEC      & Period  & P2C field   &  $K$  & $\sigma$ & $n$ & remark \\
               & (J2000)  & (J2000)  & (days)  &             &       &          &     &  \\
\hline 

 bul40\_259253 & 17.85087 & -33.1519 & 0.39951 & bul40\_2524 & 14.899  & 0.021  &  5 \\
 bul39\_15393  & 17.92076 & -30.1590 & 0.43787 & bul39\_616  & 12.685  & 0.017  &  6 & outlier \\
 bul31\_680405 & 18.04767 & -28.6609 & 0.43940 & bul32\_2167 & 14.536  & 0.011  & 11 \\
 bul45\_245148 & 18.05646 & -30.0464 & 0.44022 & bul45\_1189 & 14.410  & 0.010  &  8 \\
 bul20\_43721  & 17.98309 & -29.1750 & 0.45910 & bul20\_961  & 14.399  & 0.013  &  8 \\
 bul24\_369931 & 17.89280 & -32.9691 & 0.47306 & bul24\_1609 & 13.951  & 0.008  & 12 \\
 bul45\_256663 & 18.05978 & -30.0235 & 0.47416 & bul45\_1189 & 14.421  & 0.018  &  7 \\
 bul3\_239578  & 17.89180 & -30.3038 & 0.48308 & bul3\_791   & 14.605  & 0.082  &  4 \\
 bul22\_133009 & 17.93951 & -30.6361 & 0.48700 & bul22\_3993 & 14.331  & 0.011  &  9 \\
 bul44\_201914 & 17.82289 & -29.8616 & 0.48806 & bul44\_5324 & 11.237  & 0.006  &  6  & outlier\\
 bul20\_44361  & 17.98202 & -29.1714 & 0.48950 & bul20\_961  & 14.994  & 0.021  &  6 \\
 bul4\_207615  & 17.90976 & -30.1619 & 0.50215 & bul4\_170   & 14.405  & 0.028  &  7 \\
 bul31\_23420  & 18.03285 & -28.9437 & 0.50610 & bul31\_662  & 12.923  & 0.020  &  5  & outlier\\
 bul22\_219909 & 17.94488 & -31.1046 & 0.52411 & bul22\_815  & 12.718  & 0.007  &  6  & outlier\\
 bul39\_156007 & 17.92021 & -29.5568 & 0.52499 & bul39\_5385 & 14.119  & 0.026  &  4 \\
 bul34\_159387 & 17.96703 & -28.9742 & 0.54158 & bul34\_4631 & 14.437  & 0.017  &  7 \\
 bul40\_107313 & 17.84697 & -33.1318 & 0.54215 & bul40\_2524 & 14.430  & 0.024  &  5 \\
 bul4\_451052  & 17.91352 & -29.9600 & 0.54390 & bul4\_2323  & 14.185  & 0.013  &  8 \\
 bul5\_261524  & 17.84023 & -29.8885 & 0.55190 & bul5\_3719  & 14.320  & 0.027  &  8 \\
 bul22\_402472 & 17.94676 & -31.0980 & 0.55482 & bul22\_815  & 14.408  & 0.014  &  6 \\
 bul4\_439650  & 17.91289 & -29.9957 & 0.55603 & bul4\_2323  & 14.419  & 0.011  &  7 \\
 bul32\_87588  & 18.05029 & -28.6886 & 0.57740 & bul32\_2167 & 13.994  & 0.014  &  9 \\
 bul22\_144794 & 17.94068 & -30.6044 & 0.57910 & bul22\_3993 & 14.364  & 0.016  &  9 \\
 bul25\_187102 & 17.90468 & -33.1508 & 0.57918 & bul25\_693  & 14.244  & 0.015  &  5 \\
 bul22\_144296 & 17.94163 & -30.5937 & 0.58562 & bul22\_3993 & 14.261  & 0.012  &  9 \\
 bul43\_130470 & 17.58358 & -27.5280 & 0.59590 & bul43\_351  & 14.246  & 0.013  & 10 \\
 bul22\_333396 & 17.94312 & -30.5768 & 0.61422 & bul22\_3993 & 14.525  & 0.036  &  4 \\
 bul24\_527005 & 17.89388 & -32.9262 & 0.61678 & bul24\_1609 & 14.356  & 0.011  & 10 \\
 bul3\_250926  & 17.89106 & -30.2063 & 0.63179 & bul3\_1755  & 13.017  & 0.013  &  6  & outlier \\
 bul34\_159754 & 17.96594 & -29.0003 & 0.63539 & bul34\_4631 & 14.066  & 0.016  &  9 \\
 bul45\_245339 & 18.05951 & -30.0290 & 0.63597 & bul45\_1189 & 14.034  & 0.045  &  6 \\
 bul34\_145186 & 17.96658 & -29.0164 & 0.66322 & bul34\_4631 & 14.193  & 0.025  &  4 \\
 bul30\_604974 & 18.03144 & -29.1517 & 0.67131 & bul30\_1107 & 14.361  & 0.025  &  8 \\
 bul34\_389919 & 17.96842 & -29.0303 & 0.67304 & bul34\_4631 & 14.334  & 0.016  &  7 \\
 bul39\_15358  & 17.92118 & -30.1017 & 0.69769 & bul39\_616  & 14.323  & 0.040  &  5 \\
 bul24\_517291 & 17.89567 & -32.9539 & 0.76522 & bul24\_1609 & 14.237  & 0.010  & 13 \\
 bul4\_207388  & 17.90973 & -30.1789 & 0.78642 & bul4\_170   & 14.090  & 0.015  &  6 \\

\hline 
\end{tabular} 

\end{table*}

\section{RR Lyrae stars}

\begin{figure*}

\begin{minipage}{0.49\textwidth}
\resizebox{\hsize}{!}{\includegraphics{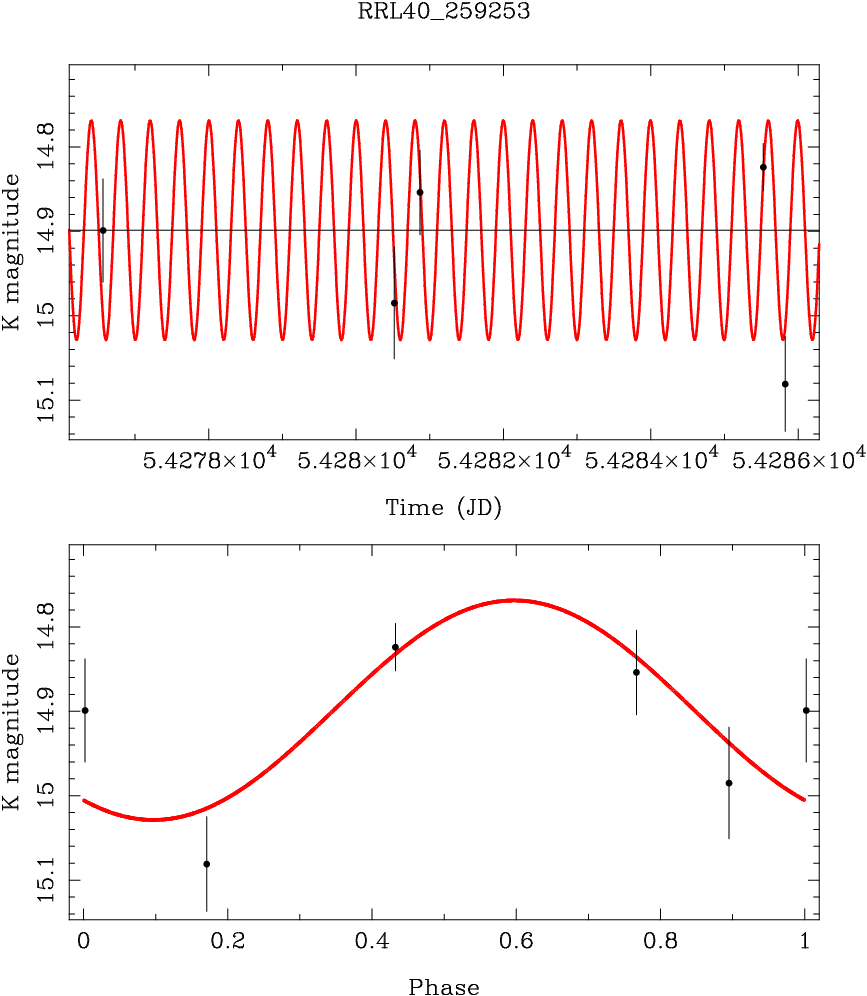}}
\end{minipage}
\begin{minipage}{0.49\textwidth}
\resizebox{\hsize}{!}{\includegraphics{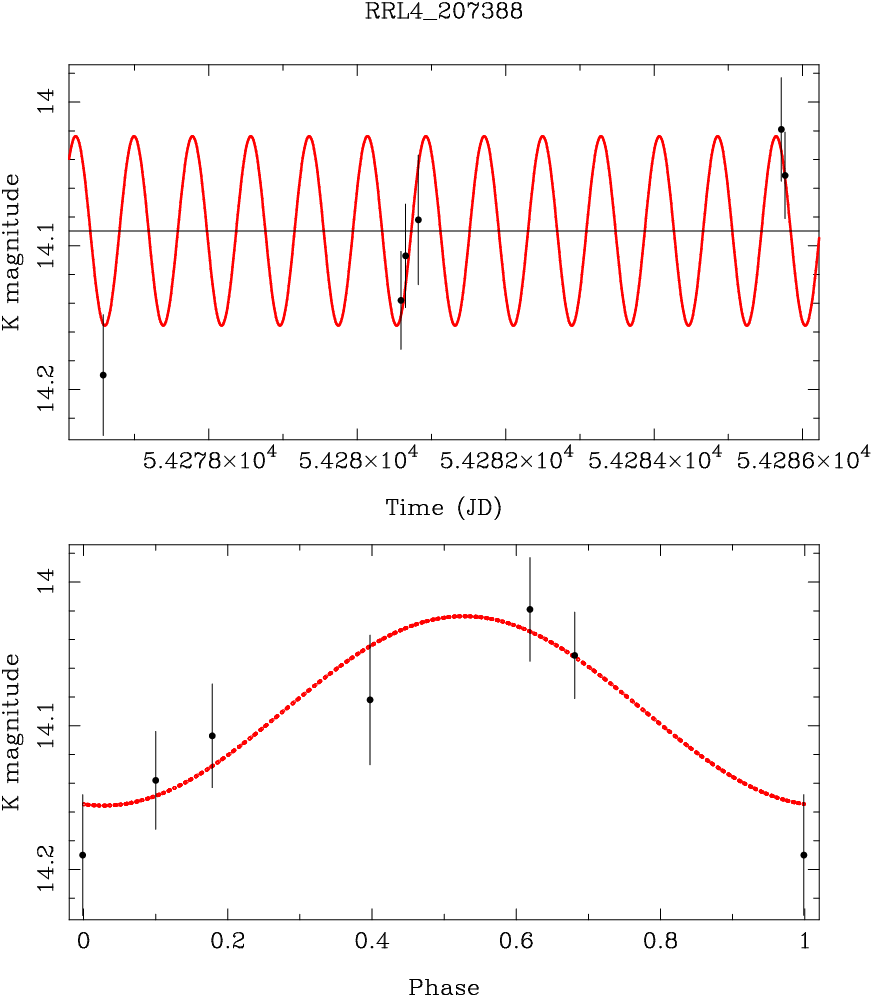}}
\end{minipage}

\caption[]{
The $K$-band light curves of two RR Lyrae.
The top panel shows the magnitude versus time, the bottom panel the phased light curve.
}
\label{Fig-KLCR}
\end{figure*}

Collinge et al. (2006) present a catalog of 1888 fundamental-mode RR
Lyrae stars in the \OG\ fields. It was verified which of them happen to
be located in the SOFI fields of view, and for those the $K$-band
light curve was extracted, and fitted with a sine-curve with the known
period.  Table~\ref{TsampleRRL} present the results, with columns 2-4
taken from Collinge et al. Column~5 lists in which P2C field it is
located and in the analysis the corresponding reddening for that field
is taken (see below). The last columns lists the mean magnitude, the
error, and the number of epochs.  Figure~\ref{Fig-KLCR} shows the
light curve for 2 stars.

\section{Analysis}

\subsection{P2C: $K$-band PL-relation}

The $K$-band magnitudes are de-reddened using the model by Marshall et al. (2006).
They present 3-dimensional $K_{\rm s}$-band extinction
along 64~000 lines of sight at a resolution of 15\arcmin\ in typically
four distance bins.  Using
VizieR\footnote{http://vizier.u-strasbg.fr/viz-bin/VizieR} the
available data within 20\arcmin\ radius of the targets were retrieved.
The mean and dispersion was determined of the $A_{\rm K}$ values in
the bin that correspond to a distance larger than 4 kpc (for 2 stars a
smaller distance had to be adopted).  Table~\ref{Tsample} lists in
Columns~7-9 the mean, dispersion and the number of data points used.

Columns~10 of this table list the $A_{\rm K}$ value derived by
multiplying by 0.12 (based on the reddening law of Cardelli et
al. 1989) the $A_{\rm V}$ value for the OGLE-{\sc ii} GB fields by Sumi (2004).

Based on the MACHO survey towards the GB Popowski et
al. (2003) derive visual extinction for 9717 elements at a resolution
of about 4\arcmin. The available data within 5\arcmin\ radius of the targets
was retrieved.  The mean and dispersion was determined, and multiplied
by 0.12 to obtain the $K$-band extinction, which is listed with the
number of elements used as Columns 11-13 in Table~\ref{Tsample}.

The comparison of the reddenings based on the Marshall et al. model
and the values from Sumi and Popowski et al. suggest that there
are no systematic effects and that the error in the adopted reddening is realistic.

\begin{figure} 
\includegraphics[width=85mm]{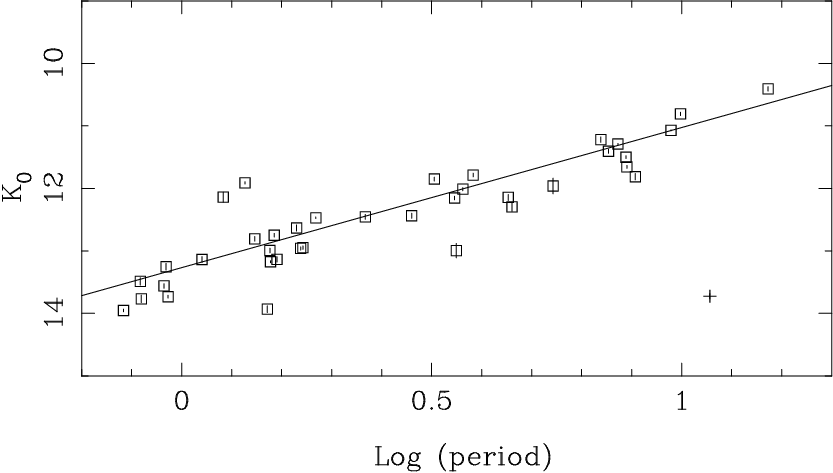} 
\caption[]{ 
The $K$-band PL-relation for P2C in the Bulge. The line is a best-fit excluding the cross.
} 
\label{Fig-PL} 
\end{figure}

The observed $K$-magnitudes are then de-reddened, and Fig.~\ref{Fig-PL}
shows for the P2C the observed period-luminosity relation in the $K$-band,
together with the best-fitting line (excluding the cross), 
$K_0 = (-2.24 \pm 0.14) (\log P - 1.2) + (10.578 \pm 0.099)$, with an rms of 0.41 mag.
This is based on a weighted least-squares fit, where the error in
$A_{\rm K}$ is added in quadrature to the error in $K$. 
Eliminating the possible three outliers near a period of 1.3 days gives
a slope of $-2.41 \pm 0.08$ and a ZP of 10.529 $\pm$ 0.059 with an rms
of 0.28 mag.

This dispersion is largely due to the intrinsic depth of the Bulge which is
of the order of 1 kpc (e.g. Babusiaux \& Gilmore 2005), and was
already seen in the dispersion around the Wesenheit $PL$-relation by KU.

Plotting the residuals versus galactic longitude and latitude
suggested a slight dependence on these parameters, and a general
linear fit was made (excluding only one object):
\begin{displaymath}
K_0 = (-2.24 \pm 0.13) (\log P - 1.2) + (10.60 \pm 0.17) 
\end{displaymath}
\begin{equation}
\hspace{20mm} +\; (-0.028 \pm 0.031)\;  l + (0.005 \pm 0.031)\; b
\end{equation}
The slope agrees within the error bar with that of the $PL$-relation derived by 
M06 which is also on the 2MASS system: 
$M_{K_0} = (-2.41 \pm 0.05) (\log P - 1.2) + (-4.00 \pm 0.02)$ with a
dispersion of 0.14 mag.
 
In order to derive the distance to the GC a Monte-Carlo simulation was performed. 
Data sets with new values for $A_{\rm K}$ and $K$ were generated based on
the mean values and (Gaussian) errors listed in Table~\ref{Tsample}, and unweighted 
fits were made for a fixed slope of $-2.41$. 
This results in a zero point (ZP) of 10.512 $\pm$ 0.013.

The DM to the GC therefore is 14.51 $\pm$ 0.02, corresponding to 7.99 $\pm$ 0.09 kpc. 
The error takes into account the precision in our ZP and the error in the
ZP of the M06 relation.

Even adding in quadrature an additional 0.015 mag to the error on the
photometry (allowing for an underestimate of the adapted error on the
transformation from instrumental magnitude to the 2MASS system), and
doubling the error on the $A_{\rm K}$ value (with a minimum error of
0.03) increases the formal error bar on the derived ZP only to 0.025 mag, and
the error on the DM to 0.03 (0.12 kpc).

%
%

\subsection{P2C: the Wesenheit-index PL-relation}

An alternative method to derive the distance is to use a
reddening-free Wesenheit index. The most appropriate one is 
$WIK= I - \alpha (I-K)$ as the $I$-band magnitudes are available from OGLE.
Unfortunately the available $I$-band data (Pritzl et al. 2003) for
stars with $K$-band data in M06 is insufficient to derive an empirical
relation, but one can use theoretical $PL$-relations as derived in several bands by Di Criscienzo (2007).
These theoretical relations are in good agreement with observations: 
The slope of the theoretical $I$-band $PL$-relation, $-2.10 \pm 0.06$, 
is in agreement with the observed one of $-2.03 \pm 0.03$ (Pritzl et al. 2003), while
In the $K$-band the theoretical slope of $-2.38 \pm 0.02$ is in good
agreement with the observed $-2.41 \pm 0.02$ in M06. For an assumed
mixing length parameter of 1.5, one needs to adopt a fiducial [Fe/H]
value of $-1.63$ in the theoretical $PL$-relation to obtain the
observed $I$ magnitude at a typical period of $\log P = 0.5$ (Pritzl
et al. 2003), and similarly a fiducial [Fe/H] of $-2.55$ to obtain the
observed $K$ magnitude at $\log P = 0.5$ (M06).

The theoretical relation is, depending on the coefficient $\alpha$ in the Wesenheit-relation (Di Criscienzo, private comm.) 
$WIK= (-1.13 \pm 0.02) - (2.53 \pm 0.04) \log P + (0.06 \pm 0.01) $[Fe/H] for $\alpha_1= 1.252$ 
(standard reddening law) or, 
$     (-1.17 \pm 0.02) - (2.57 \pm 0.04) \log P + (0.06 \pm 0.01) $[Fe/H] for $\alpha_2= 1.316$ 
(allowing for anomalous reddening in $I$, see Sumi 2004).
For an [Fe/H] value of $-2.0 \pm 0.4$ one expects ZPs of  $-1.25 \pm 0.04$ ($\alpha_1$) and $-1.29 \pm 0.04$ ($\alpha_2$)

A potential complication is that the theoretical relation is derived
for the Bessell \& Brett (1988) photometric system, while in
particular the \OG\ $I$-filter is slightly non-standard.  An
appropriate transformation is applied to both the 2MASS $K$, and \OG\
$I$ to the Bessell \& Brett system. It turns out however that for the
typical $(V-I)$ colours of the P2C in the observed fields and the
particular value of the slope in the Wesenheit-index the corrections
largely cancel, and the effect on the derived ZP is minimal.

For $\alpha_1$ the derived slope from the observations is $-2.25 \pm 0.17$, 
for $\alpha_2$ $-2.27 \pm 0.18$. The agreement between the observed and theoretical slopes is less 
good (about 1.6-1.7$\sigma$ deviation) than that in the $K$-band (only 1.3$\sigma$ deviation).

When the slope is fixed to the theoretical value one obtains ZPs of
13.19 $\pm$ 0.01 and 13.04 $\pm$ 0.01, respectively, based on a
Monte-Carlo simulation.  Depending on the reddening law adopted one
obtains a DM of about 14.44 (7.73 kpc), or 14.33 (7.35 kpc) with an
internal error bar of 0.04 mag (0.14 kpc).


\subsection{RR Lyrae}

Sollima et al. (2006) derive an empirical PLK-relation based on 15 GC
and the LMC cluster Reticulum, which for a metallicity of [Fe/H]= $-1.0$
(the average metallicity of RR Lyrae in the GB, see Walker \& Terndrup
1991), reads $M_{\rm K} = (-2.38 \pm 0.04) \log P - (1.13 \pm 0.13)$,
which was calibrated against the trigonometric parallax of RR Lyra
(Benedict et al. 2002).

\begin{figure} 
\includegraphics[width=85mm]{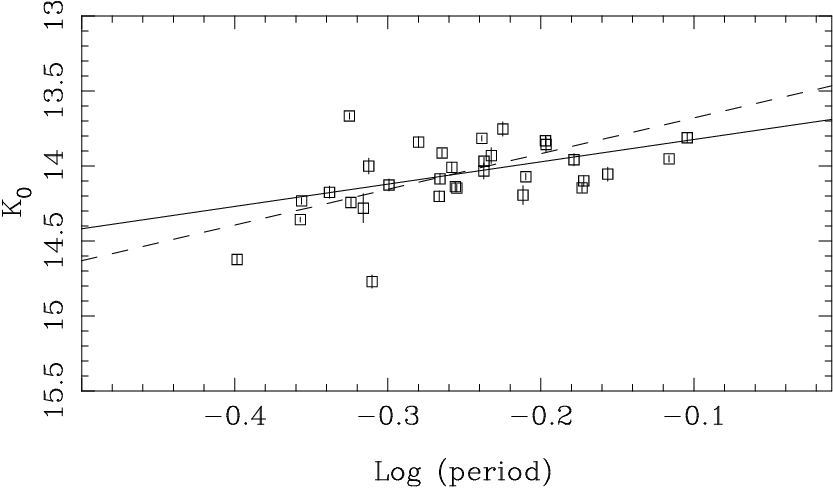} 
\caption[]{ 
The $K$-band PL-relation for the RR Lyrae stars in the Bulge.  The
line is a best-fit excluding the outliers labeled in Table~\ref{TsampleRRL} which
fall outside the plot. The dotted line is the fit for a fixed slope of $-2.38$. 
} 
\label{Fig-PLRRL} 
\end{figure}

Removing five bright outliers the PL-relation becomes (cf. Figure~\ref{Fig-PLRRL}):
\begin{displaymath}
K_0 = (-1.36 \pm 0.49) \log P + (13.63 \pm 0.17) 
\end{displaymath}
\begin{equation}
\hspace{20mm} +\; (+0.019 \pm 0.021)\;  l - (0.025 \pm 0.043)\; b
\end{equation}
The error in the derived slope is large but formally agrees within
2$\sigma$ with the empirical slope by Sollima et al.  For a fixed
slope of $-2.38$, the ZP becomes 13.39 $\pm$ 0.13 (based on a
Monte-Carlo simulation), resulting in a DM of 14.52 $\pm$ 0.18 (8.0 $\pm$ 0.7 kpc).

\section{Discussion} 

The $PL$-relation in the $K$-band of P2C in the GB is derived.  The
slope is found to be in agreement with that derived by M06.
Fixing the slope to their more accurate value implies a DM to the GC
of 14.51 with a formal error bar of 0.03.

There is also the systematic error bar to consider.
M06 presented $JHK$ period-luminosity relations based on 46 P2C with
periods between 1.2 and 80 days in 26 Galactic globular clusters (GCs). 
For the absolute magnitude scale they adopted a relation between
absolute $V$-magnitude of the Horizontal Branch and metallicity
(Gratton et al. 2003),
which in turn is calibrated using main-sequence fitting to three GCs.
This calibration implies an RR Lyrae based LMC DM of 18.50 $\pm$ 0.09 (Gratton et al. 2003).
M06 show that the DM based on P2C in the LMC and their
$K$-band $PL$-relation is also compatible with 18.5.

They also show that there is no significant trend with metallicity
over the range $-2.2 \la$ [Fe/H] $\la -0.5$, in agreement with
theoretical predictions (Bono et al. 1997, 2003, Di Criscienzo et
al. 2007), and indicating that this $K$-band $PL$-relation should be
applicable for GC P2C as well. The metallicity of the P2C in the Bulge
is unknown but that of RR Lyrae is estimated to be on average [Fe/H]= $-1.0$ 
(Walker \& Terndrup 1991).  Any difference between that metallicity
and the mean metallicity of about [Fe/H]= $-1.5$ of the GCs in M06
would result in an uncertainty in the ZP of \less 0.03 mag.

Di Criscienzo et al. (2007) show that adopting a different $M_{\rm V}$-[Fe/H] 
relation has a negligible effect on the derived slope of
the NIR $PL$-relations.

The ZP in the calibrating relation by Gratton et al. has a formal
error of 0.07 and this has to be considered as a source of
systematic uncertainty in the derived distance.

There is other source of (random) error to consider, namely how representative
this particular set of 39 stars (minus the one outlier) that defines
the $PL$-relation is in view of the fact that they scatter along the
line-of-sight due to the intrinsic depth of the Bulge. To simulate
this, additional Monte-Carlo simulations were carried out. Random
samples of 38 stars were selected from the original sample, and the
$PL$-relation re-derived. The dispersion in the ZP is about 0.11 mag. 
This is likely a slight overestimate as in this approach the
randomly drawn samples do not necessarily have the large spread in
period that the true sample was selected to have.


The DM to the GC we derive based on the P2C is 14.51 $\pm$ 0.12 (random) $\pm$ 0.07 (syst).
The random error could be improved further by observing additional
systems when the full OGLE-{\sc iii} database becomes available.

Based on the serendipitously observed RR Lyrae stars in the field a DM
of 14.52 $\pm$ 0.18 is derived. The error bar is for 50\% due to the
uncertainty in the adopted absolute magnitude of RR Lyra itself
(Sollima et al.). Their PL-relation led to an LMC distance of 18.54
$\pm$ 0.15.  If instead we would {\em assume} the LMC distance to be
18.50 (to be consistent with the P2C calibration) then we would find a
DM of 14.48 $\pm$ 0.13 (random) $\pm$ 0.07 (syst), were the systematic
error comes from the uncertainty in the ZP of the observed LMC
PL-relation. Like for the P2C sample, there is an additional 0.11 mag
systematic uncertainty due to the limited sample size. The final DM to
the GC based on the $K$-band RR Lyrae stars is 14.48 $\pm$ 0.17 (random) $\pm$ 0.07 (syst).

As the two distance estimates are derived independently, they can be
averaged and the best empirical estimate of the DM to the GC based on
the current data is 14.50 $\pm$ 0.10 (random) $\pm$ 0.07 (syst).  
%

The theoretical WIK relation gives a formal result of 14.44 (or 14.33
with anomalous reddening) with an internal error bar of 0.04 mag. A
random error of 0.11 has to be added to this, due to the limited sample,
and as the theoretical relation is tied to the observed relations of
M06 a similar systematic error bar of 0.07 has to be considered.

Although within the error bar of the purely empirical results, it
brings up the question of reddening and the reddening law. An
additional absorption in $K$ of 0.1 mag would bring all three methods
in very good agreement. On the other hand, the reddening estimates
listed in Table~1 are in excellent agreement and are based on $(J-K)$
colours (Marschall et al. 2006), $(V-I)$ (Sumi 2004), and $(V-R)$
(Popowski et al. 2003). If the absorption in $K$ were underestimated,
it would also imply an underestimate of the reddening in the other
maps, or a significantly higher selective reddening 
$A_{\rm K}/A_{\rm V} \sim 0.16$ instead of 0.12.

A final remark is that independent distances to some of these P2C may
be obtained using surface-brightness relations (e.g. Groenewegen 2004)
and the Baade-Wesselink technique. This would require better sampled
$K$-band light curves than were needed for the present study and
well-sampled radial velocity curves. Although observationally
expensive it would give an improved understanding on the systematic
error in the present analysis.

\acknowledgements{  
MG would like to thank ESO astronomers Valentin Ivanov and Alessandro
Ederoclite for their support at the telescope and discussion on the
SOFI pipeline, and Evelien Vanhollebeke for the introduction to the
DoPhot package.
The first draft of this paper was written when MG was a short-term
visitor at the Max-Planck Institut f\"ur Astrophysik, Garching.
AU was partly supported by the the Polish MNiSW grant N20303032/4275.
GB thanks PRIN/INAF 2006 (PI: F. Ferraro) for partial support.
This publication makes use of data products from the Two Micron All
Sky Survey, which is a joint project of the University of
Massachusetts and the Infrared Processing and Analysis
Center/California Institute of Technology, funded by the National
Aeronautics and Space Administration and the National Science Foundation.
%
%
}

{}


\begin{thebibliography}{} 

\bibitem[]{} Babusiaux, C., \& Gilmore, G. 2005  MNRAS, 358, 1309

\bibitem[]{} Bessell, M.S., \& Brett, J.M. 1988, PASP 100, 1134

\bibitem[]{} Benedict, G.F., McArthur, B.E., Fredrick, L., et al. 2002, ApJ, 581, 115 

\bibitem[]{} Bono, G., Caputo, F., \& Santolamazza, P. 1997, A\&A, 317, 171

\bibitem[]{} Bono, G., Caputo, F., Castellani, V., et al. 2003, MNRAS, 344, 1097 

\bibitem[]{} Cardelli, J.A., Clayton, G.C., \& Mathis, J.S. 1989, ApJ, 345, 245

\bibitem[]{} Carney, B.W., Fulbright J.P., Terndrup D.M., et al. 1995, AJ, 110, 1674

\bibitem[]{} Catchpole, R., Whitelock, P.A., Feast, M.W., et al. 1999, IAU Symposium 192, Astronomical Society of the Pacific, p. 89


\bibitem[]{} Collinge, M.J., Sumi, T., \& Fabrycky, D. 2006, ApJ, 651, 197

\bibitem[]{} Di Criscienzo, M., Caputo, F., Marconi, M.,\& Cassisi, S. 2007, A\&A 471, 893

\bibitem[]{} Eisenhauer, F., Sch\"odel, R., Genzel, R., et al. 2003, ApJ, 597, L121

\bibitem[]{} Eisenhauer, F., Genzel, R., Alexander, T., et al. 2005, ApJ, 628, 246

\bibitem[]{} Feast, M.W., \& Whitelock P.A., 1997, MNRAS, 291, 683

\bibitem[]{} Fernley, J.A., Longmore, A.J., Jameson, R.F., Watson, F.G., \& Wesselink, T. 1987, MNRAS, 226, 927

\bibitem[]{} Gratton, R.G., Bragaglia, A., Carretta, E., et al. 2003, A\&A, 408, 529

\bibitem[]{} Groenewegen, M.A.T. 2004, MNRAS, 353, 903

\bibitem[]{} Groenewegen, M.A.T., \& Blommaert, J.A.D.L. 2005, A\&A 443, 143

\bibitem[]{} Kubiak, M., \& Udalski, A. 2003, AcA, 53, 117 (KU)

\bibitem[]{} Marshall, D.J., Robin, A.C., Reyle\', C., Schultheis, M., \& Picaud, S. 2006, A\&A, 453, 635

\bibitem[]{} Matsunaga, N., Fukushi, H., \& Nakada, Y., et al. 2006, MNRAS, 370, 1979 (M06)

\bibitem[]{} McNamara, D.H., Madsen, J.B., Barnes, J., \& Ericksen B.F. 2000, PASP, 112, 202

\bibitem[]{} Nishiyama, S., Nagata, T., Sata, K., et al. 2006, ApJ, 647, 1093 

\bibitem[]{} Paczy\'nski, B., \& Stanek, C. 1998, ApJ, 494, L129

\bibitem[]{} Popowski, P., Cook, K.H., \& Becker, A.C., 2003, AJ, 126, 2910

\bibitem[]{} Pritzl, B.J., Smith, H.A., Stetson, P.B., et al. 2003, AJ 126, 1381

\bibitem[]{} Reid, M.J. 1993, ARA\&A, 31, 345

\bibitem[]{} Schechter, P.L., Mateo, M.L., \& Saha, A. 1993, PASP, 105, 1342

\bibitem[]{} Sollima, A., Cacciari, C., \& Valenti, E. 2006, MNRAS, 372, 675

\bibitem[]{} Sumi, T., 2004, MNRAS, 349, 193

\bibitem[]{} Vanhollebeke, E., Groenewegen, M.A.T., \& Girardi, L. 2008, A\&A, submitted

\bibitem[]{} Walker, A.R., \& Terndrup, D.M. 1991, ApJ, 378, 119

\bibitem[]{} Zucker, S., Alexander, T., Gillessen, S., Eisenhauer, F., \& Genzel, R., 2006, ApJ, 639, L21

\end{thebibliography}
\end{document}